\documentclass[10pt]{iopart}

\usepackage{graphicx}
\usepackage{url}

\begin{document}

\title[Validation and benchmarking of two PIC codes for a glow discharge]{Validation and benchmarking of two particle-in-cell codes for a glow discharge}

\author{Johan Carlsson$^1$\footnote{Crow Radio and Plasma Science}, Alexander Khrabrov$^1$, Igor Kaganovich$^1$, Timothy Sommerer$^2$ and David Keating$^3$}
\address{$^1$ Princeton Plasma Physics Laboratory, P.O. Box 451, Princeton, NJ 08543, USA}
\address{$^2$ General Electric Global Research, 1 Research Circle, Niskayuna, NY 12309, USA}
\address{$^3$ Department of Physics, University of California at Berkeley, Berkeley, CA 94720, USA}
\ead{carlsson@pppl.gov}

\begin{abstract}
The two particle-in-cell codes EDIPIC and LSP are benchmarked and validated for a parallel-plate glow discharge in helium, in which the axial electric field had been carefully measured, primarily to investigate and improve the fidelity of their collision models. The scattering anisotropy of electron-impact ionization, as well as the value of the secondary-electron emission yield, are not well known in this case. The experimental uncertainty for the emission yield corresponds to a factor of two variation in the cathode current. If the emission yield is tuned to make the cathode current computed by each code match the experiment, the computed electric fields are in excellent agreement with each other, and within about 10\% of the experimental value. The non-monotonic variation of the width of the cathode fall with the applied voltage seen in the experiment is reproduced by both codes. The electron temperature in the negative glow is within experimental error bars for both codes, but the density of slow trapped electrons is underestimated. A more detailed code comparison done for several synthetic cases of electron-beam injection into helium gas shows that the codes are in excellent agreement for ionization rate, as well as for elastic and excitation collisions with isotropic scattering pattern.
The remaining significant discrepancies between the two codes are due to differences in their electron binary-collision models, and for anisotropic scattering due to elastic and excitation collisions.
\end{abstract}


\maketitle
\ioptwocol

\section{Introduction\label{sec:intro}}

When performing computer simulations of complicated systems, such as low-temperature plasmas, it is typically not immediately obvious whether the results are correct, or made invalid by simplistic physics models, weak algorithms, poor numerical convergence, software implemented or input files configured incorrectly, or a combination of subtly interacting factors. To distinguish spurious simulation results from physical ones, skeptical analysis and systematic scrutiny are thus prudent.
The importance of verification, benchmarking and validation of plasma-simulation codes is therefore increasingly being recognized~\cite{Turner2013}.
An early example of such an effort for low-temperature plasmas is the pioneering work of Surendra for a capacitively-coupled radiofrequency (RF) discharge~\cite{Surendra1995}, where simulation results from twelve different codes (including four particle-in-cell codes) were compared with each other (benchmarking) and with experiment (validation).

There is an inherent conflict between benchmarking and validation. For a successful benchmarking exercise, where differences between codes are small and fully understood, it is preferable to have a simple simulation model. For a successful validation exercise, a realistic, and therefore more complex, simulation model might be necessary. In this paper we perform both validation and benchmarking. We first do validation with a fairly complete simulation model, and subsequently do additional benchmarking using much simpler simulation models to allow us to identify the source of differences between codes. Succesful verification (testing that equations are correctly solved) of main aspects of both codes had been done prior to the work reported here and will not be further discussed.

We benchmark and validate the two particle-in-cell (PIC) codes EDIPIC~\cite{Sydorenko2006} and LSP~6.95~\cite{Clark2005} for a short (or obstructed) parallel-plate glow discharge in helium at 3.5~Torr~\cite{DenHartog1988,Lawler1991}, where the anode terminates the negative-glow region. The discharge had water-cooled aluminum electrodes with 3.2~cm diameter, located 0.62~cm apart, and was operated in the moderately abnormal regime. In preliminary experiments, a segmented cathode was used consisting of an inner cylindrical segment with 1.6~cm diameter and a close-fitting outer annular segment with outer diameter of 3.2~cm. When both segments were held at the same potential, the measured current densities on each were virtually identical. It was thus concluded that radial effects were negligible, and a one-dimensional approximation was well justified, where only the distance along the axis needs to be resolved numerically in simulations.

The plasma in the device is sustained by secondary electrons emitted from the cathode surface, primarily by incident ions. The effective secondary-electron emission yield, with a value of around 0.3, is thus one of the most important physical parameters. However, as will be discussed in Section~\ref{sec:UQ} below, its value is not precisely known. The secondary electrons are rapidly accelerated in the cathode fall and create multiple electron/ion pairs through impact ionization. The electron density in the cathode-fall region is small. The negative-glow region has a quasi-neutral, cool plasma confined in a weak potential well and the electric field in this region is very weak. As stated in~\cite{DenHartog1988} and references therein, electron temperature in the negative glow is determined by a balance between thermalization via elastic collisions with neutrals and heating via Coulomb collisions with the fast electrons accelerated by the cathode fall. The main source of electrons for this relaxation process is the ionization within the negative glow by ``ballistic'' electrons originating in the cathode fall. Kinetic effects are thus important in both regions of the plasma, with the electrons from the cathode fall forming a significant tail on the electron energy distribution function in the negative glow. The location of the boundary between the cathode-fall and negative-glow regions depends on the applied voltage in a complex way. With increasing voltage, the cathode fall first widens, then narrows again. For the lowest and highest voltages, 173~V and 600~V, respectively, the cathode fall is approximately 0.38~cm wide. For the intermediate voltages, 211~V, 261~V and 356~V, respectively, the width of the cathode fall is approximately 0.28~cm.

The electric field in the glow discharge was measured using various laser diagnostics, primarily optogalvanic detection of Rydberg atoms~\cite{Lawler1980,Doughty1984a}. Optogalvanic spectroscopy can be used when the neutral-neutral collision frequency is sufficiently high for Rydberg atoms to predominantly ionize, rather than relax radiatively, the latter process being utilized by laser-induced fluoresence. Optogalvanic spectroscopy measures spikes in discharge current generated by this increased ionization due to a resonant laser pulse. The resonance frequency depends on the Stark shift of degenerate excited levels and is therefore relatively straightforwardly related to the local DC electric field. By measuring the discharge current as the laser frequency is scanned, the strength of the electric field can thus be determined, in this case with a claimed accuracy of 1\%~\cite{DenHartog1989}. The electron density and temperature in the negative-glow region were inferred from Monte-Carlo simulations using optogalvanic-spectroscopy data as input.

The one-dimensional nature of the configuration, together with the precisely measured electric field, makes this particular experiment attractive as a validation target for plasma-simulation codes, or "as a set of benchmark experiments for further modeling of the cathode region", as proposed by the authors~\cite{DenHartog1989}. Such a validation was done for the 211~V case by Parker, Hitchon and Lawler~\cite{Parker1993} using a code based on an algorithm known as the "convected scheme" for self-consistently solving the Poisson-Boltzmann equations in two-dimensional phase space (axial location and velocity)~\cite{Hitchon1989,Sommerer1989,Sommerer1991}. Excellent agreement was found for the electric field, but the electron density found by the code was lower than the value inferred from the optogalvanic-spectroscopy data. The Parker validation will be discussed in more detailed in Sections~\ref{sec:UQ} and \ref{sec:results1}.

\begin{figure}
\begin{center}
\includegraphics[width = 0.7\columnwidth, angle = -90]{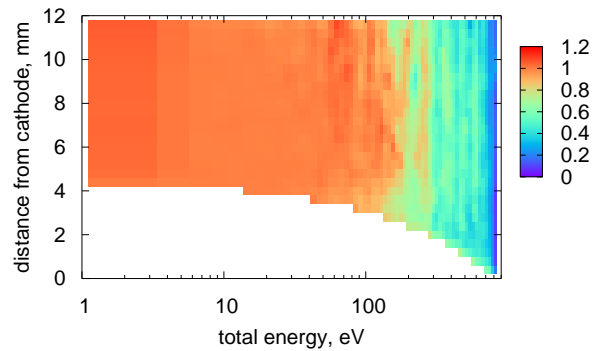}
\end{center}
\caption{Anisotropy map for a 1.2~cm glow discharge at 800~V in 3.5~Torr helium (from previous, unpublished work, in which a similar obstructed glow discharge in helium was studied)}
\label{fig:anisotropymap}
\end{figure}

It should be noted that there is also a strong theoretical motivation for choosing a cold-cathode glow discharge at several hundred volts for validation and benchmarking 
of numerical simulations. Although quite a simple system conceptually, it requires numerical treatment that must account for non-local kinetics with strongly anisotropic and/or 
non-Maxwellian velocity distributions. The system is strongly inhomogeneous with the dynamic range of electron energies and spatial densities covering about three orders of magnitude (between the cathode-fall and the negative-glow regions). Also the anisotropy has a strong variation along the axis of the device, as illustrated by Fig.~\ref{fig:anisotropymap}, which concisely presents the structure of the electron component of the plasma in the phase space by mapping the anisotropy parameter $a=\langle\frac{2}{3}\sin^2\theta\rangle$, where $\theta$ is the angle between the velocity vector and the dicharge axis, versus the full energy (kinetic plus potential) and distance from the cathode. The angle brackets $\langle...\rangle$ denote averaging over the local distribution. For a fully isotropic distribution, $a=1$, while, for example, for a distribution uniform over a hemi-sphere $a=\frac{1}{3}$ and for a beam aligned with the electric field $a=0$. The discharge in this example is at 800~V in 3.5~Torr helium, with an electrode spacing of 1.2~cm. The map indicates that electrons accelerated to high energies in the cathode fall are strongly anisotropic, with a non-local distribution approximately defined by the value of the energy integral (which corresponds to the location the electron originates at in the cathode avalanche). There is also an intermediate group (below 150~eV), relaxed in momentum but not in energy, whose energy distribution "remembers" the non-local structure. The cold (0.1~eV) electrons in the negative glow, whose density is about 1000 times higher than in the cathode sheath, are not mapped. The theory can be found, e.g., in~\cite{Kolobov1992}.
%

The remainder of this paper is organized as follows. The simulation models used for this benchmarking/validation exercise by the two PIC codes EDPIC and LSP~6.95 are described in Section~\ref{sec:model}. Uncertainty quantification for two of the key simulation-model parameters is the topic of Section~\ref{sec:UQ}. The results from the glow-discharge simulations are presented in Section~\ref{sec:results1}. To identify the cause of some code discrepancies, a set of simplified cases of electron-beam injection into helium gas are simulated, as discussed in Section~\ref{sec:results2}. A comparison between fixed-voltage and fixed-current glow-discharge simulation results is done in Section~\ref{sec:circuit}. The main findings and conclusions are presented in Section~\ref{sec:conclusions}.

\section{Elementary processes and simulation model\label{sec:model}}

The simulation model is one-dimensional, with only the axial direction (the distance between the circular electrodes) resolved. The two codes EDIPIC and LSP~6.95 that are used for the benchmarking and validation exercise presented here are both particle-in-cell (PIC) codes with Monte-Carlo collision (MCC) models. For both codes the electric field is found by solving the Poisson equation with Dirichlet-Dirichlet boundary conditions on an equidistant mesh, discretized using finite differences with second-order discretization error. A spatial resolution of 10~$\mu{}m$ is found sufficient to resolve both the Debye length at peak density in the negative glow and the anode sheath.
For both codes, the direct-implicit particle advance~\cite{Langdon1983} is used for the electrons for all the inter-code benchmark simulations. For LSP, the results are virtually identical with direct-implicit and explicit Boris particle advance, respectively.
The typical number of macro particles launched per cell of each species are in the range of 8--40, with an initial density of 1--5$\times 10^{10} / cm^3$.
Both codes are parallelized using MPI. LSP uses the same domain decomposition for fields and macroparticles. EDIPIC has a direct, serial field solve and the macroparticles in each MPI process are distributed over the whole computational domain. For a typical simulation, around a dozen processor cores can be efficiently used. The time step is limited by the cell transit time for an electron accelerated over the whole cathode fall, just over a picosecond for the high-voltage case.

As mentioned in Section~\ref{sec:intro}, ion-impact secondary-electron emission (iSEE) is a critical process for maintaining the discharge. The standard version of LSP has an iSEE model that is deterministic. When an ion macroparticle impinges on a wall, it will emit one electron macroparticle with numerical weight reduced by a factor equal to the iSEE emission yield. Using this model severly slows down the simulation by introducing ever more macroparticles with ever smaller numerical weight. This numerical inefficiency could be mitigated by merging lightweight particles. However, more seriously, we fail to achieve numerical convergence when the standard LSP iSEE model is configured to emit only every few time steps. The simulation results varies significantly and unexpectedly depending on whether the electrons are emitted every one, two or three time steps. Rather than trying to make this existing, deterministic iSEE emission model perform correctly, we implemented a new probabilistic model, similar to the one in EDIPIC. In these latter models, secondary electron macroparticles are emitted with a probability equal to the emission yield when an ion macroparticle hits a wall. For a typical simulation, with a quasineutral initial plasma and equal numbers of ion and electron macroparticles, the emitted electrons will then conveniently have the same numerical weight as the initial elecrons. The true value for the iSEE emission yield is difficult to estimate, as it tends to drift over time as the condition of the electrode surfaces changes due to plasma exposure. The parameter uncertainty of the emission yield is an important topic and is discussed in more detail in Section~\ref{sec:UQ} below.

The predominant collisions are charge exchange of ions on neutrals; elastic, excitation and ionization for electrons on neutrals, and Coulomb collisions for electrons on electrons. For each charged-neutral collision, a random atom is sampled from the background gas and used as the collision target, taking into account the neutral temperature. The charge exchange (CX) cross section varies weakly with energy over the relevant energy range and is well approximated by Eq.~(3.9.4) in the text book by Smirnov~\cite{Smirnov2015}. The CX model in LSP was found to give unexpected result and was modified. Our new version of the CX model was then verified by reproducing analytic results for simulations of ion-beam injection into neutral gas with the field solve disabled (not shown).

\begin{figure}
\begin{center}
\includegraphics[width = 0.9\columnwidth]{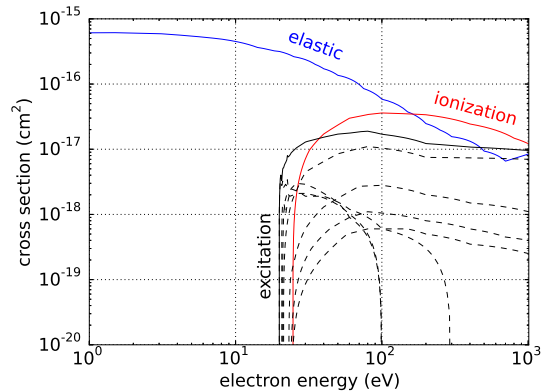}
\end{center}
\caption{Tabulated electron-neutral cross sections used in the simulations. Elastic in blue and ionization in red. The black graphs are for excitation collisions with dashed line for the seven lowest excitation levels and the solid line for a single, consolidated (energy-weighted) excitation cross section.}
\label{fig:XsectionsHe}
\end{figure}

For LSP, an arbitrary number of excitation levels can be used, and for helium a table with cross sections for excitation to the lowest seven excited states for electron energies up to one keV is included in the code suite and is plotted in Fig.~\ref{fig:XsectionsHe}.
The provenance of this data is not known to us, but it is in reasonable agreement with data from well-known sources, except at energies above 200~eV, where the LSP cross sections seem to fall off too slowly. For the current simulations, this discrepancy is believed to occur at energies too high to significantly affect the simulation results. For EDIPIC a single, consolidated excitation cross section is needed. Such a consolidated cross section was calculated by summing up the energy-weighted lowest seven individual excitation cross sections, shown as the solid, black graph in Fig.~\ref{fig:XsectionsHe}. Two LSP simulations were done to compare the results using the original and consolidated excitation cross sections, respectively. The simplified excitation model is not found to noticably increase the error bar set by other uncertainties, most notably the experimental measurement error of the cross sections.

The elastic, excitation and ionization collisions are all anisotropic. 
For elastic and excitation collisions, both codes use an anisotropy model recently developed specifically for helium~\cite{Khrabrov2012}, 
based on energy-dependent screened-Coulomb scattering~\cite{Okhrimovskyy2002, Berlenguer1999, Wentzel1927}: 
\begin{equation}
\label{eq:dcs}
\frac{1}{\sigma(E)} \frac{d\sigma}{d\Omega}(E,\theta) = \frac{1}{4\pi} \frac {1+\varepsilon} {\left(1+\varepsilon\sin^{2}\frac{\theta}{2}\right)^2},
\end{equation}
where the normalised energy $\varepsilon$ is defined as $8E/E_{aniso}(E)$ and $E_{aniso}$ is the anisotropy parameter, on the order of several atomic energy units.
For the electron-neutral ionization, the same screened-Coulomb model is applied, but with a constant (not energy-dependent) screening length treated as an adjustable parameter that determines the level of anisotropy for scattering on the given species of the target atoms.
In adopting the above approach to scattering (of both primary and secondary electrons) in ionizing collisions, we follow previous work such as Refs.~\cite{Parker1993} and \cite{Boeuf1982}, where ad-hoc approximations were employed due to lack of reliable reference data. It should be noted that eventually we found the elastic-scattering model of Ref.~\cite{Khrabrov2012}, applied at corresponding post-collision energies, to work equally well as the adjustable model with the chosen value of $E_{aniso}$. 

Another model-related structural uncertainty, or inadequacy, of the current simulations is the neglect of atoms in metastable excited states. Metastables can, in principle, introduce associative ionization as well as create secondary electrons as they diffuse to the cathode and impinge. However, at least for the low and intermediate voltages cases, the associative ionization is only about 5\% of the total ionization~\cite{DenHartog1989}, which is within the error bar of the cross section. Additionally, it would require the implementation of a rather complex model, which could at best modestly benefit the validation results, at the expense of making the code benchmarking more complicated. Metastables are therefore excluded from the simulation models used by EDIPIC and LSP here.

Unlike EDIPIC, LSP does not have null collisions and determines if a collision occurred for each macroparticle of a collisional species. To facilitate the benchmarking, EDIPIC's null-collision algorithm is disabled for the current simulations.

A subtle structural uncertainty can be introduced by pseudo-random-number generators (PRNGs) with statistical deficiences. In EDIPIC, the standard PRNG is replaced with the Mersenne Twister~\cite{Matsumoto1998}, which has emerged as the most commonly used high-quality PRNG. In LSP, for expediency, the standard PRNG is instead replaced by drand48, which is part of the Portable Operating System Interface (POSIX) standard and an intrisic function available from the standard C library of the compilers used.


\section{Parametric uncertainty quantification\label{sec:UQ}}

\begin{figure}
\begin{center}
\includegraphics[width = 0.9\columnwidth]{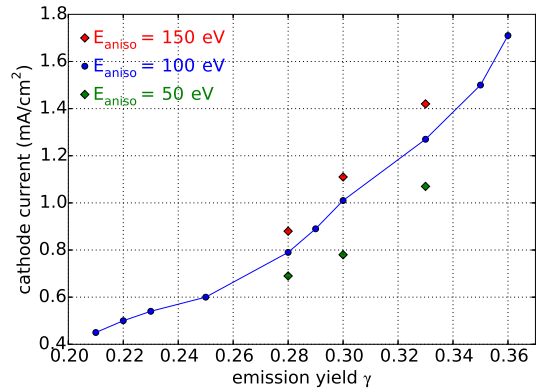}
\end{center}
\caption{Cathode current computed with LSP vs. $\gamma$ for three different values of the $E_{aniso}$ parameter in the Okhrimovskyy anisotropy model for ionization~\cite{Okhrimovskyy2002}}
\label{fig:CurrentVsGamma}
\end{figure}

Experimentally, the effective SEE emission yield $\gamma$ has a large parameter uncertainty, $\gamma \approx 0.32 \pm 0.10$. The value of the screening energy $E_{aniso}$ in the scattering model is also poorly known, but believed to be around 100~eV, based on the values of this parameter for elastic collisions at scattering energies where the ionization cross-section is near its maximum~\cite{Khrabrov2012}. To assess how sensitively the simulation results depend on the values of these two simulation-model parameters we perform a number of simulations with LSP of the highest-voltage (600~V) case. To reduce the execution time of the simulations, a relatively large time step of 5~ps was used, too large for full temporal convergence. The values of the computed cathode current are therefore slightly high, but the scaling with $\gamma$ and $E_{aniso}$ is nevertheless valid. The results are shown in Fig.~\ref{fig:CurrentVsGamma}.
As can be seen, the current grows rapidly with increasing values of $\gamma$. For example, increasing $\gamma$ by 50\% (from 0.22 to 0.33) increases the current by 154\% (from 0.50~mA/cm${}^2$ to 1.27~mA/cm${}^2$). In the simulations, $\gamma$ is therefore for practical purposes a free parameter that can be adjusted to reproduce the experimental cathode current.

Changes to $E_{aniso}$ that are well within the error bar also have large impact on the cathode current, albeit less dramatic than for $\gamma$. Also $E_{aniso}$ can thus be used as a free parameter to adjust the current. For example, ($\gamma = 0.28, E_{aniso} = 100~eV$) result in almost the same current as ($\gamma = 0.30, E_{aniso} = 50~eV$).

\section{Validation for glow discharge\label{sec:results1}}

Validation of EDIPIC and LSP is done for the glow discharge~\cite{DenHartog1989} using the simulation model described in Section~\ref{sec:model}. Three cases are simulated: the lowest voltage (173~V), one of the intermediate voltages (211~V) and the highest voltage (600~V). The 211~V case was simulated previously by Parker \emph{et al.}~\cite{Parker1993}, not with a PIC-MCC code, but with a code that solves the Poisson-Boltzmann equations using the so-called "convected scheme"~\cite{Hitchon1989}.

For both the important parameters $\gamma$ and $E_{aniso}$, the experimental error bars are too large to offer any real guidance on which values to use in the simulations. We therefore, somewhat arbitrarily, choose $E_{aniso} = 100~eV$ for all the simulations and use $\gamma$ as a free parameter that is adjusted in increments of 0.01 to reproduce the cathode current reported from the experiment.

\begin{figure}
\begin{center}
\includegraphics[width = 0.9\columnwidth]{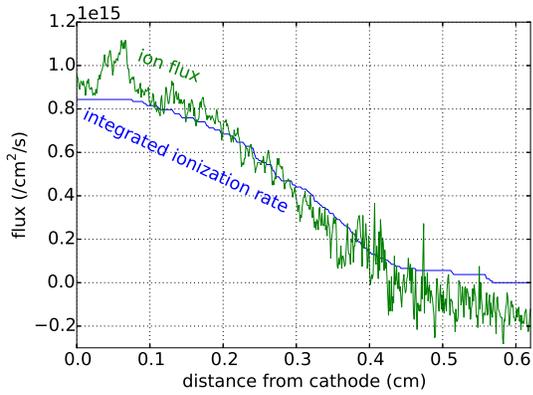}
\end{center}
\caption{Integral of ionization rate (blue graph) and ion flux (green graph) at end of EDIPIC simulation of 173~V case. Approximate overlap is indicative of near steady-state solution, except in negative-glow region ($>$~0.35~cm).}
\label{fig:SteadyState}
\end{figure}

\begin{figure}
\begin{center}
\includegraphics[width = 0.9\columnwidth]{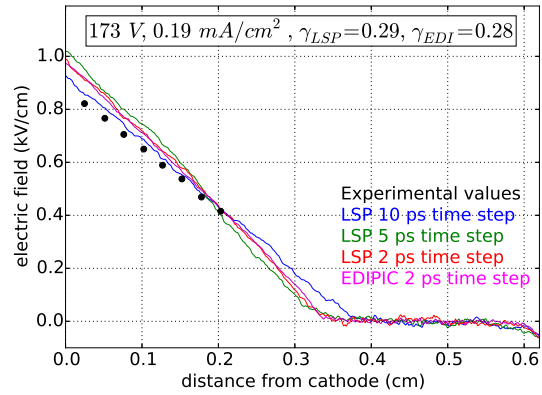}
\end{center}
\caption{Electric field for the 173~V case. Black circles are experimental values, blue, green and red graphs are LSP results with decreasing time steps (10, 5 and 2~ps, respectively). The magenta graph is the EDIPIC result with 2~ps time step.}
\label{fig:ElectricField173V}
\end{figure}

For the 173~V case, EDIPIC reproduces the experimental cathode current of 0.19~mA/cm${}^2$ with $\gamma = 0.28$, whereas LSP gets closest to the experimental current for $\gamma = 0.29$. For LSP, temporal convergence is achieved with a time step of 2~ps. With 1~ps time step, the early simulation results are virtually identical as with double the time step, and are therfore not ran to steady state. The limitation on the time step seems to be imposed by the cell transit time for an electron accelerated over the full cathode fall. Steady state is reached after 120~$\mu{}s$, or 60~million time steps, as shown in Fig.~\ref{fig:SteadyState}.
It should be noted that this steady state does not apply to the accumulation of cold electrons in the negative glow, which occurs on a millisecond time scale and will be discussed in more detail below. Electric-field profiles from these simulations, as well as from the experiment (data digitized from Fig.~4 of Ref.~\cite{DenHartog1989}), are shown in Fig.~\ref{fig:ElectricField173V}.
As can be seen, when the time step is small enough for temporal convergence, the two codes are in excellent agreement (red and magenta graphs). Compared to experiment, the codes predict an electric field that is about 10\% larger at the cathode and decreases faster than linearly away from it.

\begin{figure}
\begin{center}
\includegraphics[width = 0.9\columnwidth]{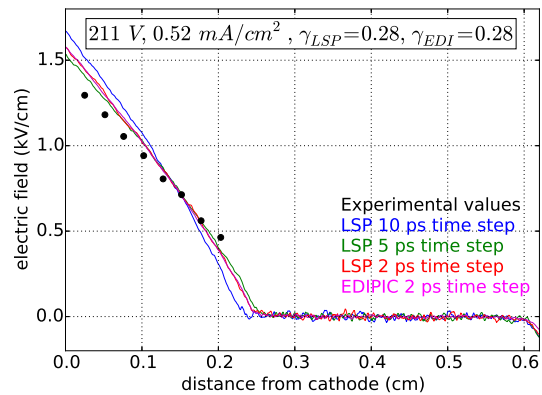}
\end{center}
\caption{Electric field for the 211~V case. Black circles are experimental values, blue, green and red graphs are LSP results with decreasing time steps (10, 5 and 2~ps, respectively). The magenta graph is the EDIPIC result with 2~ps time step.}
\label{fig:ElectricField211V}
\end{figure}

For the 211~V case, both codes reproduce the experimental cathode current of 0.52~mA/cm${}^2$ with $\gamma = 0.28$. Like for the lower voltage, the 2~ps time step is sufficient for temporal convergence and 120~$\mu{}s$ for steady state. The results are shown in Fig.~\ref{fig:ElectricField211V}.
As for the lower-voltage case, the two codes are in excellent agreement (red and magenta graphs), but overestimate the electric field at the cathode by about 15\%.

\begin{figure}
\begin{center}
\includegraphics[width = 0.9\columnwidth]{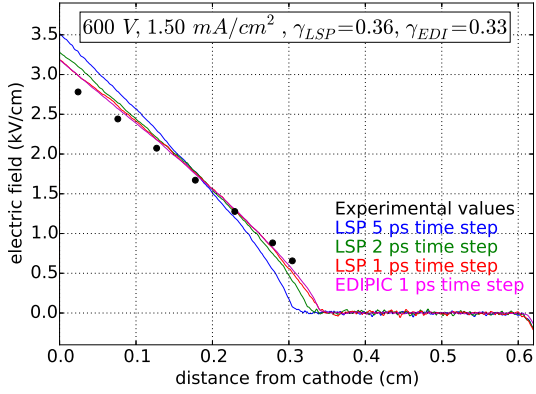}
\end{center}
\caption{Electric field for the 600~V case. Black circles are experimental values, blue, green and red graphs are LSP results with decreasing time steps (5, 2 and 1~ps, respectively). The magenta graph is the EDIPIC result with 1~ps time step.}
\label{fig:ElectricField600V}
\end{figure}

For the 600~V case, EDIPIC requires $\gamma = 0.33$ to match the experimental cathode current of 1.50~mA/cm${}^2$, and LSP needs $\gamma = 0.36$. To reach temporal convergence, the time step has to be halved to 1~ps. The electric-field profile is shown in Fig.~\ref{fig:ElectricField600V}.
As for the two lower-voltage cases, the codes agree (red and magenta graphs), but show better agreement with experiment, with the electric field at the cathode about 7\% too large.

The main results from the glow-discharge validation are summarized in Tab.~\ref{tab:validation}.
\begin{table}
\begin{center}
\begin{tabular}{|l|c|c|c|}
\hline
 & 173~V & 211~V & 600~V \\
\hline
\hline
Exp. $\gamma$ & $\approx 0.32$ & $\approx 0.32$ & $\approx 0.32$ \\
\hline
Exp. $E^0 (kV/cm)$ & 0.90 & 1.43 & 3.02 \\
\hline
Exp. $d_c \, (cm)$ & 0.38 & 0.30 & 0.40 \\
\hline
\hline
EDIPIC $\gamma$ & 0.28 & 0.28 & 0.33 \\
\hline
EDIPIC $E^0 (kV/cm)$ & 0.97 & 1.58 & 3.18 \\
\hline
EDIPIC $d_c \, (cm)$ & 0.35 & 0.25 & 0.34 \\
\hline
\hline
LSP $\gamma$ & 0.29 & 0.28 & 0.36 \\
\hline
LSP $E^0 (kV/cm)$ & 0.99 & 1.57 & 3.18 \\
\hline
LSP $d_c \, (cm)$ & 0.35 & 0.25 & 0.34 \\
\hline
\hline
\end{tabular}
\end{center}
\caption{For 173~V the current density is 0.19~mA/cm${}^2$, for 211~V it is 0.52~mA/cm${}^2$ and for 600~V it is 1.50~mA/cm${}^2$. In the simulations, the emission yield $\gamma$ was adjusted in each case to reproduce the experimental current density.
$E^0$ is the axial electric field at the cathode surface. $d_c$ is the width of the cathode-fall region.
Experimental values from Tab.~1 of Ref.~\cite{DenHartog1988}.}
\label{tab:validation}
\end{table}
As a benchmarking exercise, the 173~V and 211~V cases are unqualified successes, but for the 600~V case a 10\% larger $\gamma$ was needed for LSP to compute the same current. As a validation exercise, the simulations are partially succesful. Both codes correctly reproduced the dependence of the cathode-fall width on voltage, with a minimum width at intermediate voltage. In the experiment the cathode-fall width was extrapolated from the electric-field measurements to 0.38~cm at 173~V, 0.30~cm at 211~V and 0.40~cm at 600~V. In the simulations we get 0.35~cm at 173~V, 0.24~cm at 211~V and 0.34~cm at 600~V. To compensate for the shorter cathode fall the electric-field values in the simulations are larger than in the experiment, by 7\%--15\% at the cathode surface.

\begin{figure}
\begin{center}
\includegraphics[width = 0.9\columnwidth]{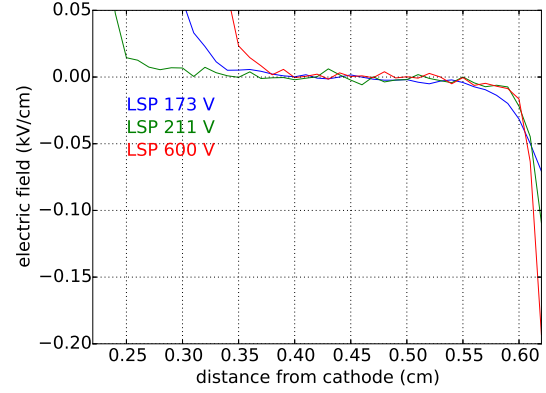}
\end{center}
\caption{LSP simulated averaged electric fields in the negative-glow region for 173~V (blue graph), 211~V (green graph) and 600~V (red graph)}
\label{fig:TimeAveragedElectricField}
\end{figure}

As can be seen in Figs.~\ref{fig:ElectricField173V}, \ref{fig:ElectricField211V} and \ref{fig:ElectricField600V}, the electric field in the negative-glow region is weak. To suppress statistical noise in the simulation data, temporal and spatial averaging is applied to the numerically converged LSP cases for each voltage. The resulting electric fields are shown in Fig.~\ref{fig:TimeAveragedElectricField}.
As can be seen, even after averaging, the computed electric field is noisy in the negative glow. However, on the reasonable assumption that the true electric field decreases monotonically from cathode to anode, one can infer that the field-reversal point is well inside the negative-glow region, by 0.10--0.15~cm, or somewhere between 0.40--0.53~cm from the cathode. The corresponding shallow potential well in the negative glow, which confines the cold electrons, does not provide a more exact field-reversal location. The depth of the potential well is of the order of a volt and it has a very wide and flat minimum (not shown).

The earlier simulations by Parker of the 211~V case using the "convected scheme"~\cite{Parker1993} showed excellent agreement with the experimentally measured electric field and found a cathode-fall width of 0.27~cm for $\gamma = 0.25$ and with a simplified model of ionization anisotropy. These earlier simulations did include the associative ionization of metastables, but did not include electron emission due to the same.

All simulations (EDIPIC, LSP and convected scheme) correctly compute an electron temperature of between 0.1~eV and 0.2~eV inside the negative glow, but underestimate the density. The reported peak density from the experiment is about $5 \times 10^{11} / cm^3$. The EDIPIC result is about $2 \times 10^{11} / cm^3$, LSP less than $10^{11} / cm^3$ and convected scheme in the range of $1-5 \times 10^{11} / cm^3$ (the exact value was not reported).

The cold-electron accumulation in the negative glow is primarily
determined by trapping of electrons produced in ionization events
within the plasma potential well, and of some of those
accelerated in the cathode fall (which produce the
ionization). Correctly simulating this process requires an accurate
model of binary electron-electron Coulomb collisions as well as
milliseconds of physical simulation time. The relevant (or at least, the longest one involved) time scale for establishing the density profile of cold plasma in the negative glow can be roughly estimated as the ambipolar
diffusion time. Based on $T_e=0.1~eV$, helium ion mobility $\mu = 10~cm^2/(V s)$ and the length scale of 0.1~cm, this equilibration time is on the order of 10~ms. The EDIPIC binary-collsion model has recently been thoroughly validated as part of yet-unpublished efforts to simulate runaway electrons and breakdown and we are fairly confident in its accuracy. The procedure employed in EDIPIC is the Langevin approximation based on the Fokker-Planck kinetic equation, as developed by~\cite{Manheimer1997}: the effect of Coulomb collisions on simulation particles is represented as a drift-diffusion stochastic differential equation. However, even in very long (billion-time-step) simulations, it seems unlikely that EDIPIC would reproduce the very high density values reported from the experiment. 

In the LSP simulations the density in the negative glow is approximately 2.5 times less than in the EDIPIC simulations. A likely culprit is the LSP binary-collision model, which we have neither scrutinized nor tested. However, an inadequate binary-collision model is unlikely to explain the need for a larger effective SEE emission yield in LSP for the 600~V case. To try to identify the cause of this discrepancy between the codes, we perform a number of synthetic code benchmarks, some of which are presented in the next section.

\section{Synthetic code benchmark for electron-beam injection\label{sec:results2}}

\begin{figure}
\begin{center}
\includegraphics[width = 0.9\columnwidth]{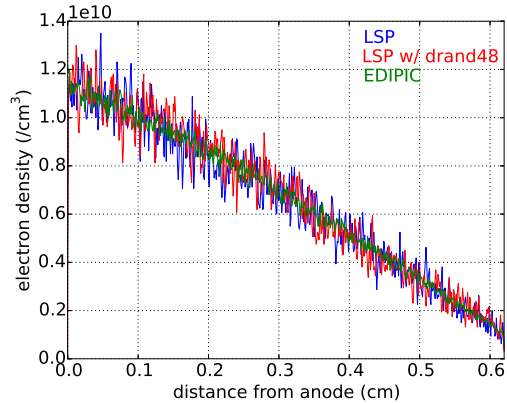}
\end{center}
\caption{100~eV electrons injected from the right boundary producing ionization with forward-scattering only. Blue graph is steady-state electron density for LSP with the standard random-number generator, red is with drand48 random numbers and green graph is with EDIPIC.}
\label{fig:beam-100eV-forward-ionization-only}
\end{figure}

To facilitate the benchmarking specifically of the collision models in the two codes, simplified simulations are ran of injection of a mono-energetic electron beam into helium gas. These synthetic-benchmark simulations are using the same set up as for the 211~V discharge, but without initial plasma and with the electric-field solve and secondary emission disabled. Fig.~\ref{fig:beam-100eV-forward-ionization-only} shows the steady-state density profiles with a 100~mA/cm${}^2$ beam of 100~eV electrons injected from the right boundary and interacting with the helium gas only via artificially anisotropic (forward-scattering only) ionization collisions.
All other collision models are temporarily disabled. Both codes are run with the same number of macroparticles, but the LSP results have more statistical noise because EDIPIC does time averaging over multiple time slices. Taking this into account, there is no statistically significant discrepancy between the results.

\begin{figure}
\begin{center}
\includegraphics[width = 0.9\columnwidth]{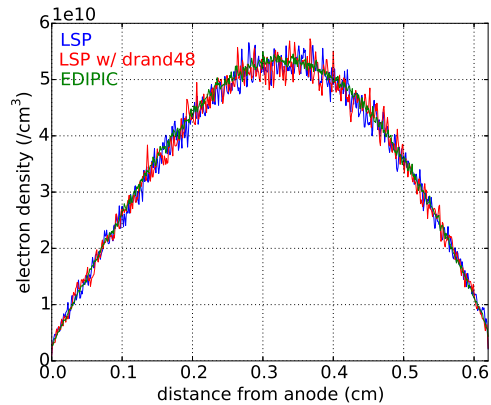}
\end{center}
\caption{100~eV electrons injected from the right boundary undergoing isotropic elastic and excitation collisions only. Blue graph is steady-state electron density for LSP with the standard random-number generator, red is with drand48 random numbers and green graph is with EDIPIC.}
\label{fig:beam-100eV-excit+elast-only-iso}
\end{figure}

Similar simulations are run with only isotropic elastic and excitation collisions with results shown in Fig.~\ref{fig:beam-100eV-excit+elast-only-iso}.
As with forward-scattering ionization only, simulations with isotropic elastic and excitation collisions only produce results that differ only in statistical noise for the two codes.

\begin{figure}
\begin{center}
\includegraphics[width = 0.9\columnwidth]{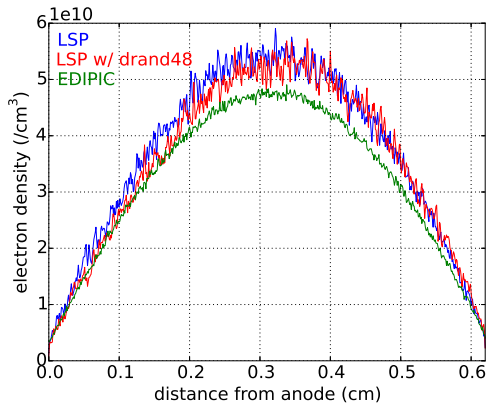}
\end{center}
\caption{100~eV electrons injected from the right boundary undergoing anisotropic (Okhrimovskyy model with $E_{aniso}$ = 100~eV) elastic and excitation collisions only. Blue graph is steady-state electron density for LSP with the standard random-number generator, red is with drand48 random numbers and green graph is with EDIPIC.}
\label{fig:beam-100eV-excit+elast-only-okh100}
\end{figure}

However, when the elastic and excitation collisions are made anisotropic the results markedly differ, as can be seen in Fig.~\ref{fig:beam-100eV-excit+elast-only-okh100}.
In this case, the Okhrimovskyy model with $E_{aniso}$ = 100~eV is used for the elastic and excitation collisions. The steady-state density is significantly lower for the EDIPIC simulation (green graph) than for the LSP simulations (blue and red graphs). There also is some difference between the LSP simulations with different random-number generators, with drand48 producing results in slightly better agreement with EDIPIC. Similar discrepancies for different random-number generators for both codes also in the more realistic glow-discharge validation simulations led us to abandon the existing random numbers in each code in favor of more modern, higher-quality ones (drand48 in LSP and the Mersenne Twister in EDIPIC).

The increased density in LSP compared to EDIPIC in the presence of anisotropic collisions leads us to hypothesize that LSP spuriously partially isotropizes the electron velocity distribution. This would be consistent with the need to increase the SEE yield for the 600~V case to compensate. This hypothesis needs further testing to be confirmed. It is also possible that spurious, numerical velocity-space diffusion plays a role~\cite{Turner2006}.

\section{Code benchmark with fixed current\label{sec:circuit}}

\begin{figure}
\begin{center}
\includegraphics[width = 0.9\columnwidth]{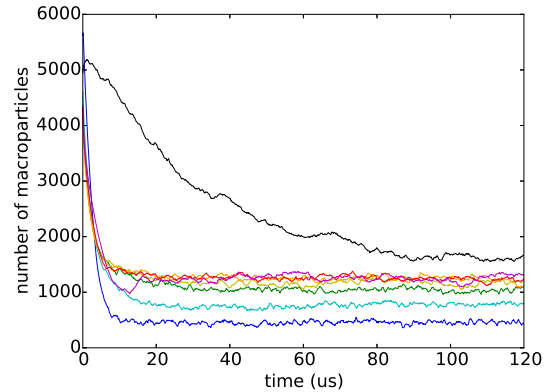}
\end{center}
\caption{Number of electron macro particles vs.~time for various LSP simulation of the 173~V / 190~$\mu{}A/cm^2$ case.
Black graph is for fixed voltage.
Blue, cyan and green graphs are for fixed current with external parallel capacitances of 50, 100 and 200~fF/cm${}^2$, respectively.
The yellow graph is for 500~fF/cm${}^2$, close to the neglected sheath capacitance.
The orange, red and magenta graphs are for 1, 2 and 5~pF/cm${}^2$, respectively.}
\label{fig:MacrosVsTime}
\end{figure}

The characteristics of the power supply used in the experiment are not known to us. For the glow-discharge simulations presented so far, the power supply is assumed to be an ideal voltage source. To assess the importance of non-zero internal resistance of the power supply, simulations are performed with both LSP and EDIPIC using their external-circuit models with a voltage source connected to the anode in series with an external resistor and the cathode connected to ground. The value of the internal resistance is made large enough for the power supply to approximate a current source. The circuit model in EDIPIC~\cite{Sydorenko2014} is based on the standard algorithm~\cite{Verboncoeur1993}, but it additionally incorporates the implicitness of surface charge deposited by electrons advanced in time using the direct implicit scheme~\cite{Friedman1981}. LSP has a less sophisticated circuit model that is found to be numerically unstable. The fixed-voltage LSP simulation that produced the Fig.~\ref{fig:ElectricField173V} 2-ps-time-step case is modified to keep the current density fixed at 190~$\mu{}A/cm^2$, but with the original LSP circuit model all electron macroparticles were lost within the first 100~ns. When the source code for the circuit model was scrutinized, it was found that all surface charge deposited by the plasma on the wall was immediately forced through the circuit model, implicitly neglecting the possibility of capacitive effects. On the hypothesis that the instability was due to neglect of the sheath capacitance, an external capacitor is added in parallel to the circuit model to compensate. The neglected sheath capacitance is estimated from the steady-state profiles of potential and charge density for the fixed-voltage case to about 540~fF/cm${}^2$. Fig.~\ref{fig:MacrosVsTime} shows the time evolution of the number of electron macro particles, which is proportional to the volume integral of the electron density, in fixed-current LSP simulations for different values of external parallel capacitance.
For increasing values of external parallel capacitance, the macro-particle number at steady state initially increases, but then levels off for a capacitance equal to or greater than the actual sheath capacitance with the macro-particle number lower than for fixed voltage (see Fig.~\ref{fig:MacrosVsTime}). The reason for the reduced electron density for fixed current compared to fixed voltage was found to be due to additional heating in the former case. For fixed current, the voltage oscillates at a frequency determined by the time constant for the RC circuit formed by plasma plus external circuit. This type of oscillation has been observed experimentally and explained theoretically in the parameter regime corresponding to the transition from a Townsend to a normal glow discharge~\cite{Kaganovich1994b,Melekhin1984}. With the true sheath capacitance, the RC oscillation frequency is about 300~kHz for the case we are simulating. The oscillation is collisionally damped through electron-electron collisions that heat the electrons. With the increase in thermal velocity, the electrons can transport current across the negative glow at reduced density.

\begin{figure}
\begin{center}
\includegraphics[width = 0.9\columnwidth]{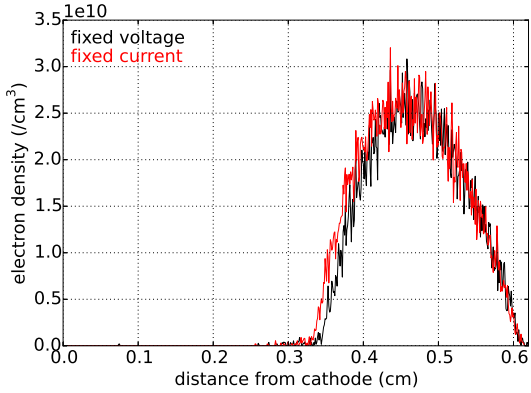}
\end{center}
\caption{Comparison of electron densities in fixed-voltage (black graph) and fixed-current (red graph) EDIPIC simulations}
\label{fig:DensFixedCurrentVsVoltage}
\end{figure}

However, in a fixed-current simulation with EDIPIC, the electron density at steady state is essentially the same as for the fixed-voltage case, as shown in Fig.~\ref{fig:DensFixedCurrentVsVoltage}.
In Fig.~\ref{fig:DensFixedCurrentVsVoltage}, the boundary between cathode fall and negative glow is closer to the cathode for fixed current than for fixed voltage. By comparing the electron-density profiles at other time slices, one finds that the boundary location is immobile for fixed voltage, but oscillates for fixed current around the same average location. The oscillation of the boundary location is consistent with a potential oscillation in the negative glow. Like for the fixed-current LSP simulations, the oscillation frequency is given by the RC time constant and the amplitude is also similar, about $\pm$10\%. Since no density reduction is associated with the voltage oscillation in EDIPIC, we conclude that the heating found in the LSP simulations is spurious and possibly caused by electron-electron collisionality being too large. We note the critical importance of code benchmarking in this case. If scrutinized in isolation, the LSP simulations results might have been difficult to identify as spurious.

\section{Conclusions\label{sec:conclusions}}

The two PIC-MCC codes EDIPIC and LSP~6.95 are benchmarked and validated. When charged-neutral collisions are isotropic and electron-electron Coulomb collisions are negligible, the two codes are in excellent agreement.

A short glow discharge~\cite{DenHartog1989} is chosen as the validation target because i) data is available from accurate electric-field measurements, ii) it is a simple configuration that can be accurately modeled as one-dimensional, and iii) it provides a very discerning test case for collision models.
Validation of the two codes for this configuration is mostly successful. The dependence of the cathode-fall width on the applied voltage, with a minimum at the intermediate voltage, is reproduced in the simulations. The axial electric field in the simulations is 7--15\% too high at the cathode surface and drops off faster than in the experiment away from the cathode. A possible explanation is the model inadequacy introduced by neglecting associative ionization by metastables in the simulations. The simulated electron temperatures in the negative glow are correct within experimental error bars, but the densities are low. Also an earlier simulation of the intermediate-voltage case using the convected scheme reported a density lower~\cite{Parker1993} than the one claimed for the experiment.

The larger values of the secondary-electron emission yield needed for the highest-voltage simulations might indicate a contribution from metastables in the experiment.
Also ultraviolet photons and fast neutrals could play a role that warrants further investigation.

The primary lesson from the work presented here is that it can be critically important to use state-of-the-art and well-tested collision models. Also, many common generators of pseudo random numbers are insufficient for simulations with Monte-Carlo collisions and can lead to subtly incorrect results. The Mersenne Twister~\cite{Matsumoto1998} provides a superior alternative and many implementations are freely available to be incorporated into PIC-MCC codes.

Simulation of low-temperature plasmas is a non-trivial endeavor. Multiple physics models for collisions, wall interation and/or chemical processes are typically invoked and each model can have large parameter uncertainty. There is also often model uncertainty, i.e. it might be unknown beforehand what physical processes are essential and should be included in the simulation model. Inadequate physics models and algorithms can cause simulation results that might not be obviously unphysical, while still being subtly wrong. Code benchmarking can help indentify such cases.
The work presented here should be seen as partial progress and part of an ongoing process to develop a set of representative benchmarking/validation problems and a more rigorous approach to simulation of low-temperature plasmas, with the goal of achieving quantitative results and ultimately a reliable predictive capability. Other groups have performed benchmarking and validation for RF discharges, including the pioneering work by Surendra~\cite{Surendra1995} and more recent work by Turner~\cite{Turner2013}.
Similar efforts should be spent on other representative types of low-temperature plasmas and benchmarking/validation of all codes should become standard practice to make simulation a more reliable and useful tool in our field.


\ack

The information, data, or work presented herein was funded in part by the Advanced Research Projects Agency-Energy (ARPA-E), U.S. Department of Energy, under Award Number DE-AR0000298.
The information, data, or work presented herein was funded in part by the Advanced Research Projects Agency-Energy (ARPA-E), U.S. Department of Energy, under Award Number DE-AR00000670.
This work was made possible by funding from the Department of Energy for the Summer Undergraduate Laboratory Internship (SULI) program. This work is supported by the US DOE Contract No.~DE-AC02-09CH11466.\\

The digital data for this paper can be found at \url{http://arks.princeton.edu/ark:/88435/dsp01x920g025r}\\

\bibliographystyle{iopart-num}
\bibliography{ltp-refs}

\providecommand{\newblock}{}
\begin{thebibliography}{10}
\expandafter\ifx\csname url\endcsname\relax
  \def\url#1{{\tt #1}}\fi
\expandafter\ifx\csname urlprefix\endcsname\relax\def\urlprefix{URL }\fi
\providecommand{\eprint}[2][]{\url{#2}}

\bibitem{Turner2013}
Turner M~M, Derzsi A, Donko Z, Eremin D, Kelly S, Lafleur T and Mussenbrock T
  2013 {\em Physics of Plasmas\/} {\bf 20} 013507

\bibitem{Surendra1995}
Surendra M 1995 {\em Plasma Sources Science and Technology\/} {\bf 4} 56

\bibitem{Sydorenko2006}
Sydorenko D 2006 {\em Particle-in-cell simulations of electron dynamics in low
  pressure discharges with magnetic fields\/} Ph.D. thesis University of
  Saskatchewan Saskatoon

\bibitem{Clark2005}
Clark R and Hughes T 2005 {\em Mission Research Corporation, Santa Barbara,
  CA\/}

\bibitem{DenHartog1988}
Den~Hartog E~A, Doughty D~A and Lawler J~E 1988 {\em Physical Review A\/} {\bf
  38}(5) 2471--2491

\bibitem{Lawler1991}
Lawler J~E, Den~Hartog E~A and Hitchon W~N~G 1991 {\em Physical Review A\/}
  {\bf 43}(8) 4427--4437

\bibitem{Lawler1980}
Lawler J~E 1980 {\em Physical Review A\/} {\bf 22} 1025

\bibitem{Doughty1984a}
Doughty D and Lawler J 1984 {\em Applied Physics Letters\/} {\bf 45} 611--613

\bibitem{DenHartog1989}
DenHartog E, O'Brian T and Lawler J 1989 {\em Physical Review Letters\/} {\bf
  62} 1500

\bibitem{Parker1993}
Parker G, Hitchon W and Lawler J 1993 {\em Physics Letters A\/} {\bf 174}
  308--312

\bibitem{Hitchon1989}
Hitchon W, Koch D and Adams J 1989 {\em Journal of Computational Physics\/}
  {\bf 83} 79--95

\bibitem{Sommerer1989}
Sommerer T, Hitchon W and Lawler J 1989 {\em Physical review letters\/} {\bf
  63} 2361

\bibitem{Sommerer1991}
Sommerer T, Hitchon W, Harvey R and Lawler J 1991 {\em Physical Review A\/}
  {\bf 43} 4452

\bibitem{Kolobov1992}
Kolobov V and Tsendin L 1992 {\em Physical Review A\/} {\bf 46} 7837

\bibitem{Langdon1983}
Langdon A~B, Cohen B~I and Friedman A 1983 {\em Journal of Computational
  Physics\/} {\bf 51} 107--138

\bibitem{Smirnov2015}
Smirnov B~M 2015 {\em Theory of gas discharge plasma\/} (Springer)

\bibitem{Khrabrov2012}
Khrabrov A~V and Kaganovich I~D 2012 {\em Physics of Plasmas\/} {\bf 19} 093511

\bibitem{Okhrimovskyy2002}
Okhrimovskyy A, Bogaerts A and Gijbels R 2002 {\em Physical Review E\/} {\bf
  65} 037402

\bibitem{Berlenguer1999}
Belenguer P and Pitchford L 1999 {\em Journal of Applied Physics\/} {\bf 86}
  4780

\bibitem{Wentzel1927}
Wentzel G 1927 {\em Z. Phys.\/} {\bf 40} 590

\bibitem{Boeuf1982}
Boeuf J and Marode E 1982 {\em Journal of Physics D: Applied Physics\/} {\bf
  15} 2169

\bibitem{Matsumoto1998}
Matsumoto M and Nishimura T 1998 {\em ACM Trans. Model. Comput. Simul.\/} {\bf
  8} 3--30 ISSN 1049-3301

\bibitem{Manheimer1997}
Manheimer W~M, Lampe M and Joyce G 1997 {\em Journal of Computational
  Physics\/} {\bf 138} 563--584

\bibitem{Turner2006}
Turner M~M 2006 {\em Physics of Plasmas\/} {\bf 13} 033506

\bibitem{Sydorenko2014}
Sydorenko D personal communication 2015

\bibitem{Verboncoeur1993}
Verboncoeur J~P, Alves M~V, Vahedi V and Birdsall C~K 1993 {\em Journal of
  Computational Physics\/} {\bf 104} 321--328

\bibitem{Friedman1981}
Friedman A, Langdon A and Cohen B 1981 {\em Comments on Plasma Physics and
  Controlled Fusion\/} {\bf 6} 225

\bibitem{Kaganovich1994b}
Kaganovich I, Fedotov M and Tsendin L 1994 {\em Sov. Phys.--Tech. Phys\/} {\bf
  39} 241--246

\bibitem{Melekhin1984}
Melekhin V and Naumov N~Y 1984 {\em Sov. Phys.--Tech. Phys\/} {\bf 29} 888--892

\end{thebibliography}

\vspace{\stretch{1}}

\noindent%
The Definitive Work version of this manuscript is published in Plasma Sources Science and Technology,\newline
Volume 26, Number 1, p.~014003 (2017), available at:\newline
\url{http://dx.doi.org/10.1088/0963-0252/26/1/014003}

\end{document}